\documentclass[aps,prl,twocolumn,nofootinbib,superscriptaddress, nobibnotes]{revtex4-1} 
\usepackage{graphicx}
\usepackage{dcolumn} 
\usepackage{bm}      
\usepackage{mathtools}
\usepackage{color}
\usepackage{bbm}
\usepackage{amssymb,amsmath,amsthm,amsfonts}
\usepackage{natbib}
\usepackage{tikz}
\usetikzlibrary{patterns}
\usepackage{hyperref}
\usepackage{float}
\makeatletter
\let\newfloat\newfloat@ltx
\makeatother
\usepackage{algorithm}
\usepackage{algorithmic}

\makeatletter
\def\blfootnote{\xdef\@thefnmark{}\@footnotetext}
\makeatother

\begin{document}

\blfootnote{This manuscript has been authored by UT-Battelle, LLC, under Contract No. DE-AC0500OR22725 with the U.S. Department of Energy. The United States Government retains and the publisher, by accepting the article for publication, acknowledges that the United States Government retains a non-exclusive, paid-up, irrevocable, worldwide license to publish or reproduce the published form of this manuscript, or allow others to do so, for the United States Government purposes. The Department of Energy will provide public access to these results of federally sponsored research in accordance with the DOE Public Access Plan.}

\title{Topological Characterization with a Twist, Condensation, and Reflection}

\author{Tushar Pandey}
\email{tusharp@tamu.edu}
\affiliation{Department of Mathematics, Texas A\&M University, College Station, TX 77840}

\author{Eugene Dumitrescu}
\email{dumitrescuef@ornl.gov}
\affiliation{Computational Sciences and Engineering Division, Oak Ridge National Laboratory, Oak Ridge, TN 37831}

\begin{abstract}

Despite its putative robustness, the realization of and control over topological quantum matter is an ongoing grand challenge. Looking forward, robust characterization protocols are needed to first certify topological substrates before they are utilized in quantum algorithms. We contribute to this grand challenge by providing a series of experimentally accessible near- and medium-term protocols assessing the fidelity of logical processes. To do so we examine logical operators and anyonic quasiparticle excitations in twisted $\mathbb{Z}_{N=2,4}$ gauge theories. Extending the finite twist, a promising route to Ising computing in it own right, to a non-contractible twist fuses prior logical operators together and results in a twisted qubit code. The code is notable for a doubled and tripled code distance for logical $Y$ and $X$ errors respectively.
Next, we review the deconfinement properties of a $\mathbb{Z}_4$ double semion condensation and provide an error correction algorithm. Based on this understanding we then present a $\mathbb{Z}_4$ topological quasiparticle reflectometry and scattering protocol. The protocol infers the topological properties of the system and serves as a high-level metric for the performance and lifetime of the interfaced topological codes. Our logical and scattering protocols are suitable for near-term devices where many physical qubits encode few logical qubits. The topological lifetime of a particle within a condensate conjugacy class, previously considered in fabricated and hetero-structured condensed-matter experiments, serves as a unifying performance metric across synthetic, qubit-based and naturally occurring topological order. 

\end{abstract}

\maketitle

\emph{Introduction --} Ref.~\citenum{Bluvstein2022} recently entangled 13 atom-based data qubits in a surface code\cite{Bravyi98} and 16 data qubits on a toric code\cite{Kitaev2003} while a superconducting qubit experiment\cite{Satzinger2021} verified the surface code's topological data via quasiparticle self- and mutual-exchange statistics. The rapidity of these developments, and the great effort required to re-scale error rates across orders of magnitude, motivate us to carefully examine the stability of topological systems with and without periodic boundary conditions. 

Verifying the salient target features of quantum systems is required before applications can be demonstrated. Once the noise characteristics are understood more complex controls, as required for the particular functionality, can be executed. Away from the topological domain, Hamiltonian learning is an increasingly active area of research where a sparse Hamiltonian representation is estimated for a physical system after a sequence of experiments. In an analogous vein, topological learning workflows are required to infer the topological quantum field theory's data from an experiment\cite{Satzinger2021}. Many microscopic distinct Hamiltonians correspond to a given topological phase, making topological learning simpler and different. This difference lies in terms of the acquisition of fundamentally different topological data describing the exchanges statistics and anyonic fusion tensors corresponding to the topological quantum field theory. 

Condensed-matter experiments previously probed these fusion relations, searching for a zero-bias electronic resonances between tunneling electrons and Majorana end modes\cite{Castelvecchi2021} and topological superconducting condensate-mediated quasiparticle (Andreev) scattering events\cite{Blonder1982, Beenakker1997, Law2009}. New devices and more exotic topological order are likewise classified in terms of topological scattering processes\cite{Gul2022}.

To explore topological symmetry breaking and scattering in synthetic quantum matter we first briefly overview the $\mathbb{Z}_2$ toric code and the twist deformation, revealing the logical twisted qubit code. We then review the $\mathbb{Z}_4$ generalization and twisted-condensation into the double Semion phase. We derive the error correcting properties for the DS phase and then discuss the consequences for spatially restricted condensation. This illustrates how quasiparticle reflection lifetimes serve as a natural metric for topological quantum error correcting architectures. 

\emph{$D(\mathbb{Z}_2)$ Toric Code -- }
\label{sec:Z_2} A toric code, based on the quantum double of $\mathbb{Z}_2\equiv \{ \pm1 \}$ ($D(\mathbb{Z}_2)$), is defined with qubits at each \textit{edge} of a two-dimensional lattice with periodic boundary conditions. Qubit $j$'s $\mathbb{C}^2$ Hilbert space is acted on by Pauli operators $\{ \sigma^X_j, \sigma^Z_j \} \equiv \{ X_j, Z_j \} $ with computational basis group action $X_j = |0\rangle_j \langle 1 |_j  + |1\rangle_j \langle 0 |_j$ and $\mathbb{Z}_2$ irreducible representation $Z_j = |0 \rangle_j \langle 0 |_j - |1\rangle_j \langle 1 |_j$. 

\begin{figure}[t!]
    \includegraphics[width = 0.5 \textwidth]{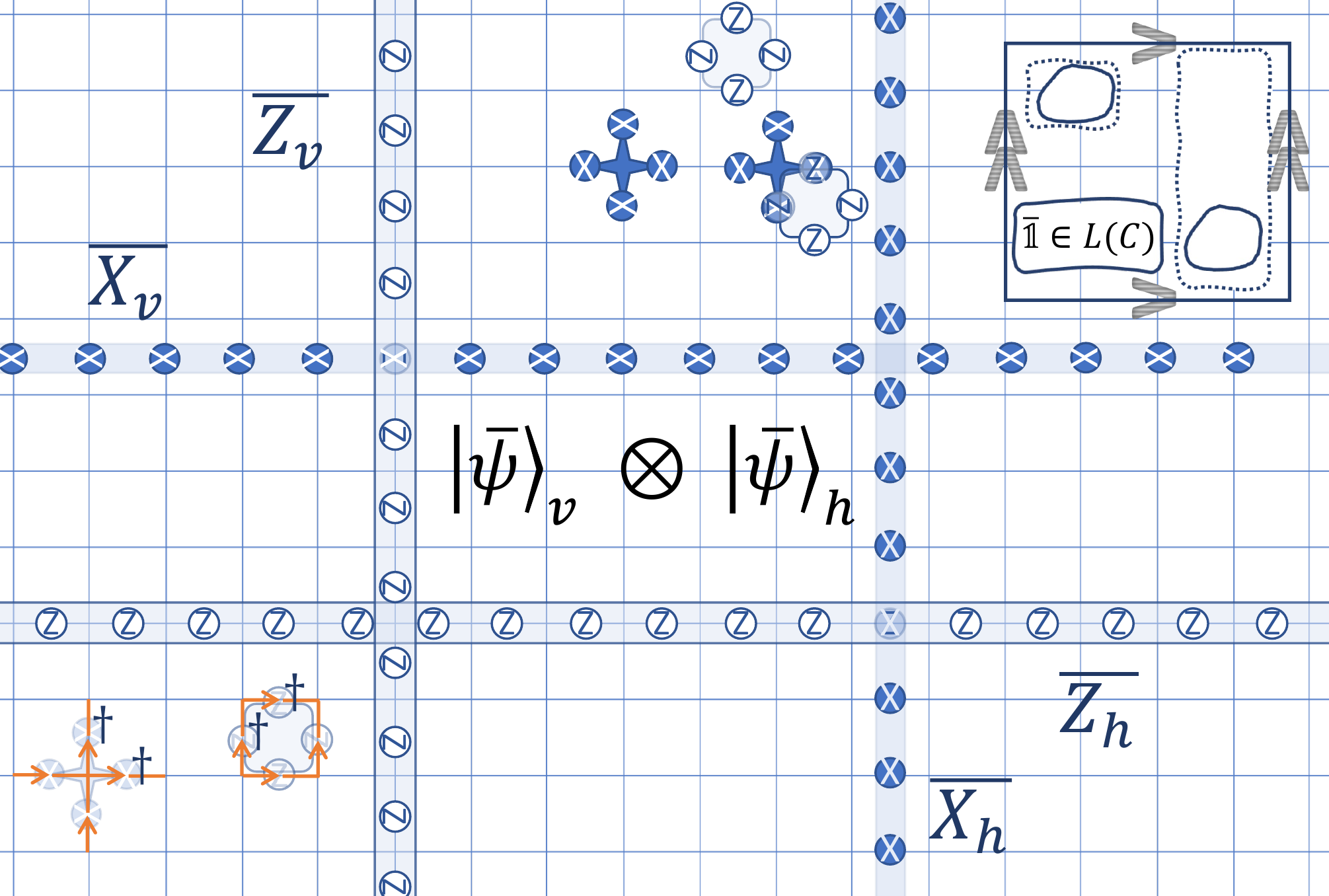}    
    \caption{Non-trivial logical operators $\in L(C)$: $\bar{X}_{h(v)},\bar{Z}_{h(v)}$ consist of transversal action of a local $(X_i,) Z_i$ on a handle (co-) cycle of \textit{horizontal} (\textit{vertical}) qubits on logical codespace $|\bar{\psi} \rangle_{v}\otimes |\bar{\psi} \rangle_{h}$. Periodic boundary conditions and examples of logical identity operators appear in a lattice-free representation in the inset. Solid (dashed) lines indicate the action of $Z(X)$ along lattice co-cycles  The $X,Z$ symbols on vertical edges are rotated for clarity. Top right inset: arrows denote boundary conditions and instances of trivial, contractible, logical operators are presented.}
    \label{fig:TC}
\end{figure}

As illustrated in Fig.~\ref{fig:TC}, define Pauli tensor product operators $A_v$ and $B_p$ supported on qubits North (N), East (E), South (S), and West (W) of a vertex $v$ or plaquette $p$ respectively:
\begin{subequations}
\label{eq:TC_generators}
\begin{eqnarray}
    A_v &= X_{N_v} \otimes X_{E_v} \otimes X_{S_v} \otimes X_{W_v}  \\ 
    B_p &= Z_{N_p} \otimes Z_{E_p} \otimes Z_{S_p} \otimes Z_{W_p}.
\end{eqnarray}
\end{subequations}
$A_v$ and $B_p$ are mutually commuting Hermitian operators with eigenvalues $\pm 1$. Summing over all commuting vertex and plaquette interactions defines the Hamiltonian 
\begin{equation}
\label{eq:Toric_code_hamiltonian}
    H = - \sum_v A_v -\sum_p B_p. 
\end{equation}

The commuting group $G = \langle A_s, B_p \rangle$ in turn defines the codespace $\mathcal{C}=\{ |\psi\rangle ; g |\psi\rangle = |\psi\rangle\ \forall g \in G \}$. The stabilizers could be tiled in two dimensions with open boundary conditions\cite{Bravyi98}, toric boundary conditions\cite{Kitaev2003}, higher handled objects, or punctured, but orientable manifolds.

One horizontal edge spin and one vertical edge spin are contained in the unit cell so that a periodic lattice with $L_x$ rows and $L_y$ columns contains $2 L_x L_y$ qubits. Each qubit is equipped with a local $\mathbb{C}^2$ Hilbert space so the dimension of the composite Hilbert space is $\mathbb{C}^{2\otimes2 L_x L_y}$. There are $L_x L_y$ plaquette, and an equal number of vertex, order two constraints. Periodic boundary conditions render two constraints redundant due to global vertex $\prod_v A_v=1$ and plaquette $\prod_p B_p = 1$ constraints.  The resulting code space, of dim$(C) = 2^{2 L_x L_y}/2^{2L_x L_y - 2} = 4$, therefore represents a pair of logical qubits, with logical-algebra $L(\mathcal{C})$ containing $\bar{X}_{v(h)} \bar{Z}_{v(h)} \in L(\mathcal{C})$. Fig.~\ref{fig:TC} illustrates instances of logical operators realized by the microscopic transversal action of local $X_i, Z_i$ along non-contractible loops of vertical or horizontal edge qubits.

\begin{figure*}
    \includegraphics[width = 0.9 \textwidth]{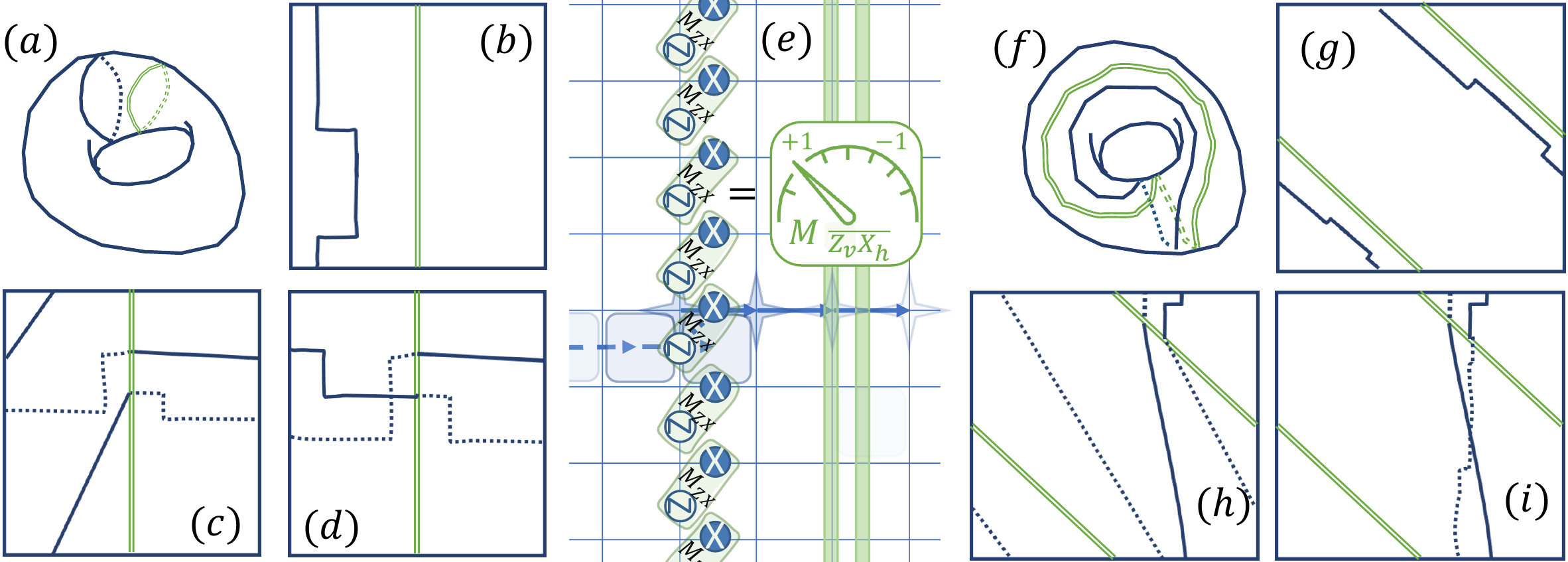}   
    \caption{Twisted the Toric code along a single handle (a-d) or traversing both handles with an additional physical twist (f-i). 
   $\bar{Z}_{t}$ is represented by the $\bar{Z}_\parallel$ operators in panels (b,g). A $X-Z$ composite logical operator perpendicular to the twist can have a no crossing $\bar{X}_t$ (c, h) or a crossing, $+i\bar{Y}_t$ (c, h)
   In the microscopic view of twist (e) measurement of $Z_v X_h$ stabilizers on a strip of qubits along a handle induces a logical twist projecting onto a logical $\bar{XZ}$ stabilizer. The paths of $e$ along the cycles between stars (solid lines) and $m$ moving along the co-cycle plaquettes (dashed lines) are exchanged at the twist (dotted line). }
    \label{fig:TC_twist}
\end{figure*}

The twist is a promising topological defect \cite{Bombin2010,Yoder2017, Krishna2020,Landahl2021} which, like surface punctures\cite{Brown2017}, increases the topologically encoded codespace. The twist generalizes simple boundary defects; action of $Z$ along lattice cycles and $X$ along co-cycles connect particles to the vacuum depending on the boundary type. That is $e \rightarrow \mathbb{I} (e), m \rightarrow m (\mathbb{I})$ at rough (smooth) boundaries\cite{Bravyi98}. However, the twist constraint $C_{t} = \langle Z_v \otimes X_h\rangle =+1$, for example induced by a series of local Bell basis measurements, mixes the cycles and co-cycles as $e \leftrightarrow m$. Fig.~\ref{fig:TC_twist} panel (e) illustrates a flux quasiparticle moving left to right by the action of $X_i$ along a horizontal co-cycle, i.e. acting on vertical edge qubits (denoted by dashed arrows), until it reaches and anti-commutes with the vertical qubit $Z$ of a measured $ZX$ stabilizer. Also, anti-commuting with the horizontal $X$ qubit of the stabilizer is the action of a string of $Z$ transporting a charge particle along the lattice cycles (denoted by solid arrows).
The logical operators in the presence of a series of twists are suitable for expressing qubit-based \cite{Krishna2020} and fermionic \cite{Landahl2021} perspectives of the Clifford gate set.

\emph{A Bell basis twist --}
\label{ss:NCT} 
The maximally entangled Bell states $\sqrt{2}| \Phi^\pm \rangle=|00\rangle \pm |11\rangle ,\sqrt{2}|\Psi^\pm\rangle=|01\rangle \pm |10\rangle$ span the two qubit Hilbert space and can be specified by a stabilizer group of signed elements $\{X_1 X_2, Y_1 Y_2, Z_1 Z_2\}$ with signs $\{ \Phi^+: +-+, \Phi^-: -++, \Psi^+: ++-, \Psi^-: --- \}$. Instead of providing \textit{two} stabilizer constraints, to select a \textit{single} Bell state, a \textit{single} stabilizer constraint admits a \textit{two-dimensional manifold}. Measuring the logical expectation value $\langle \bar{Z}_v \bar{X}_h \rangle = +1$ (if $-1$, apply recovery map) is illustrated in the center panel of Fig.~\ref{fig:TC_twist} and can be transversely implemented by measuring a \textit{local} $ZX$ expectation value (half of a Bell basis measurement which can, for example, be measured with by ancilla qubits residing on the dual lattice. 

As described above, the non contractible twist $M_{ZX}$ mixes all cycles and co-cycles passing through it. Consequentially, prior logical operators are twisted into new logical operators defined with respect to the twists' orientation. For example, in panel (a-b) we twist along a vertical handle and in panel (f-g) twist along both handles where the logical stabilizer, $[Z_v X_h] = [\mathbb{I}]$ with $[\cdot]$ denoting a logical equivalence class, is denoted by the doubled green lines and the twisted logical codespace $\mathcal{C}_t$ is spanned by $ | \bar{0} \rangle_t \equiv |\bar{0} \rangle_v |\bar{+}\rangle_h ; \;\;  | \bar{1} \rangle_t \equiv |\bar{1} \rangle_v | \bar{-} \rangle_h$. Oriented parallel with the twist, the $\bar{X}_h$ and $\bar{Z}_v$ both act as a logically equivalent, i.e. $[\bar{X}_h]=[\bar{Z}_v]=[\bar{Z}_t]$, twisted phase operator $\bar{Z}_t$. $L(\mathcal{C}_t)$ further contains $[\bar{Z}_h \bar{X}_v] = [\bar{X}_t]$ in panels (c,h) which is a logical $\bar{X}_t$ operator itself twisted around the torus handle with two non-overlapping string types. Likewise panels (d,i) illustrate  $\bar{Y}_t$, $[\bar{X}_h \bar{Y}_v] = [\bar{Y}_t]$, in panels (d,i). Solid lines are always taken to cross over dotted-lines (i.e. $Z_j X_j = +i Y_j$). Thus, at the expense of a logical qubit, the twist doubles (triples) the code distance for logical bit-phase (bit) flip errors $\bar{Y}_t (\bar{X}_t)$. 

\emph{$D(\mathbb{Z}_4)$ review --}
\label{sec:Z4}
A $\mathbb{Z}_4$ generalization, with four dimensional local Hilbert spaces is natural, appealing, and embed-able since the target ququart $\mathbb{C}^4$ may be formed by two (neighboring) $\mathbb{C}^2 \times \mathbb{C}^2$ qubits. Abusing notation, we now re-define all operators in terms of $\mathbb{Z}_4$ $X$ cyclic-shift action and $Z=\text{diag}(1,i,-1,-i)$ representing the four irreducible roots of unity 
\begin{equation} 
\label{Z_4 TC pauli operator}
    X = \sum_{k \in \mathbb Z_4} |k+1\rangle\langle k| \;\;\;\;\;\; Z = \sum_{k \in \mathbb Z_4} \omega^k |k\rangle\langle k|
\end{equation}
with $\omega = e^{2 \pi i / 4}$ and $X Z = \omega Z X $ expressing Weyl's formulation of the canonical commutation relations. Prior order two stabilizers split into order four generators,
\begin{subequations}
\label{eq:Z4_generators}
\begin{eqnarray}
    A_v &= X^\dagger_{N_v} \otimes X^\dagger_{E_v} \otimes X_{S_v} \otimes X_{W_v}^  \\ 
    B_p &= Z^\dagger_{N_p} \otimes Z_{E_p} \otimes Z_{S_p} \otimes Z^\dagger_{W_p}
\end{eqnarray}
\end{subequations}
where the adjoint action is on the N and E (N and W) qubits of a vertex (plaquette). Denoting Hermitian conjugate by $H.c.$,  Eq.~\ref{eq:Toric_code_hamiltonian} generalizes in $\mathbb{Z}_4$ as
\begin{equation} 
\label{eq:Z_4 Hamiltonian}
    \mathcal{H} = -\left[ \sum_{v}A_v + \sum_p B_p + H.c. \right]
\end{equation}

The unit cell contains a vertical and horizontal ququart, each with its $\mathbb{C}^4$ Hilbert spaces. Given $L_x$ rows and $L_y$ columns we have $2 L_x L_y$ ququarts, $L_x L_y$ order four $A_v$ constraints, and an equal number of plaquette  constraints. Like before, we remove the trivial $\prod_v A_v = 1, \prod_p B_p = 1$ constraints to arrive at a $4^{2L_x L_y}/4^{2 L_x L_y - 2} = 4^2$ dimensional logical codespace.

\begin{figure*}[t!]
    \includegraphics[width = 1.0 \textwidth]{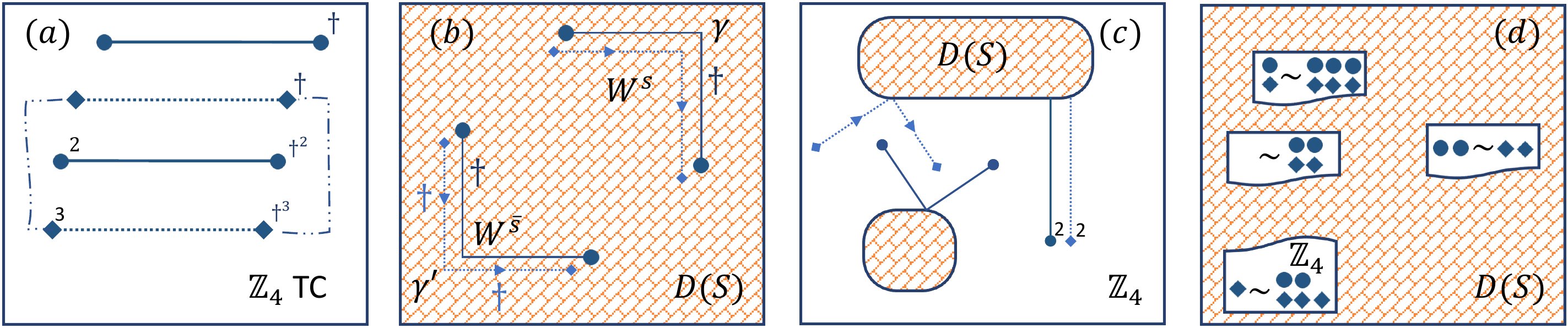}    
    \caption{(a) De-confined anyons in the TC from top to bottom: i) product of $Z$ along the lattice cycle (solid line) generates charge excitations (circles),  ii) product of $X$ along the dual lattice co-cycles (dashed lines) generating fluxes (diamonds), iii) is generated by $Z^2$ along cycles, and iv) is generated by $X^\dag$. (b) Ribbon operator of the (anti-) semion $W^{s(\bar{s})}$. (c) Along the boundary $\partial D(S)$ we have the boundary association $e^2m^2 \rightarrow \mathbb{I}$. An effect is that condensate emits pairs of charges and fluxes into the toric code in analogy with the superconducting proximity effect. Odd topological charge particles are confined in and reflected from $D(S)$ condensed regions. (d) The composite charge-flux conjugacy classes remain good quantum numbers and are discernible in small regions of un-condensed $\mathbb{Z}_4$ toric code.} 
    \label{fig:condense_and_scatter}
\end{figure*}

The 16 admissible charge-flux composite excitations  $e^a m^b$ are given by $\{ e, e^2, e^3, \mathbb{I}\}\times\{m, m^2, m^3, \mathbb{I} \}$. Transversal application of $Z_i^a,X_i^b$ along the torus' handles yields $\{ \bar{X},\bar{X}^2,\bar{X}^3=\bar{X}^\dagger, \bar{X}^4=\mathbb{I} \}$ and $ \{ \bar{Z},\bar{Z}^2,\bar{Z}^3=\bar{Z}^\dagger, \bar{Z}^4=\mathbb{I} \}$, the generators for $\bar{\mathfrak{su}}(4) \otimes \bar{\mathfrak{su}}(4) \equiv L(\mathcal{C})$ being a basis for the ququarts' logical algebra.  Fig.~\ref{fig:condense_and_scatter} (a) illustrates quasiparticles: i) charge-anti-charge pair connected with a solid line to indicate the past action of $Z$, ii) flux-anti-flux pair ($m,m^\dagger=m^3$ diamonds connected by a dotted line), iii) a $e^2-e^2$ pair, and iv) another anti-flux-flux pair. $m \times m^3 = \mathbb{I}$ so we recover the flux sector of the vacuum by fusing the particles along homologically trivial paths as denoted by the dashed-dotted lines.

\begin{figure}[t!]
    \includegraphics[width = 0.5 \textwidth]{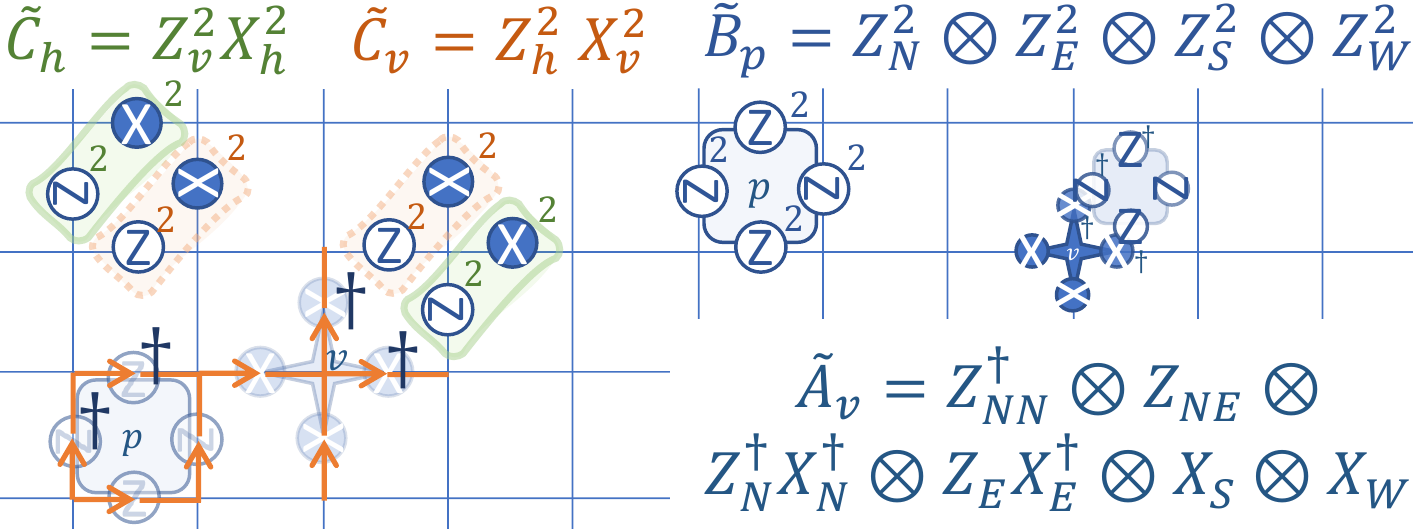}    
    \caption{$\mathbb{Z}_4$ presentation of Double Semion stabilizers. Toric stabilizers of  Eq.~\ref{eq:Z4_generators} in bottom left are translucent to indicate their expulsion from $\tilde{G}$.}
    \label{fig:DS_stab}
\end{figure}

\emph{Condensation to $D(S)$ --}
Ref.~\citenum{Ellison2022} illustrated Pauli stabilizer representations of twisted quantum doubles on composite dimensional systems, with for example global condensation of the $D(\mathbb{Z}_4)$'s $e^2 m^2$ bosonic excitation mapping onto the celebrated Double Semion (DS) phase\cite{Levin2005}. This is achieved by adding a pair of order-two constraints $\tilde{C}_{h} = X^2_{h} Z^2_{v}$ and $\tilde{C}_{v} = Z^2_{h} X^2_{v}$ as illustrated in Fig.~\ref{fig:DS_stab}. The $\tilde{C}$ stabilizers anti-commute with and remove the prior generators of Eq.~\ref{eq:Z4_generators} to form a new stabilizer group $\tilde{G} = \langle \tilde{A} , \tilde{B}, \tilde{C} \rangle$. Here $(A_v \times B_p, B^2_p)\rightarrow(\tilde{A}_v,\tilde{B}_p)$ are compatible \textit{products} of original generators as seen in Fig.~\ref{fig:DS_stab}.

In addition to condensing the codespace, topological symmetry breaking also modifies the excitations\cite{Burnell2018}. Recall $D(\mathbb{Z}_4)$'s 16 excitations and their Pauli generators. Fig~\ref{fig:condense_and_scatter} panel (a) illustrates a few examples in a lattice free representation. From top to bottom: i) a charge anti-charge ($e,e^3$) pair connected with a solid line, ii) a flux anti-flux ($m,m^3$) pair connected with a dashed line, iii) an ($e^2, e^{\dagger 2}$) pair, and iv) an ($m^3,m$) pair. Lines ii) and iv) are fused, denoted by the dashed-dotted line, to recover the null flux sector. Excitations i), ii), and iv) are confined in regions where $\langle \tilde{C}\rangle=1$.

Commuting with  $\tilde{C}$ are the short string operators $\{ X_i Z, X^{\dagger}_i Z, X_i Z^{\dagger}, X^{\dagger}_i Z^{\dagger} \}$, which are labeled into the set $\{ W_{i=h}^s, W_{i=v}^s, W_{i=h}^{\bar{s}} , W_{i=v}^{\bar{s}} \}$, and whose composition along a path $\gamma^{'}$, as in Fig.~\ref{fig:condense_and_scatter} panel (b), defines the world-lines of a deconfined semion $s$ (with ribbon operator $W_\gamma^s$) and an anti-semion $\bar{s}$ ($W_{\gamma}^{\bar{s}}$)\cite{Ellison2022}. Not illustrated is the bosonic  semion-anti-semion composite particle $s\bar{s}$ with ribbon operator $W_\gamma^{s \bar{s}}=\prod_{j \in \gamma} Z_j^2$ (equivalently $X^2$) \cite{Ellison2022}. Since all particles are self-dual non-contractible closed semion loops operators, around the two handles of the torus, constitute a logical qubit. The second qubit is controlled by (spanned on) the anti-semion's algebra (Hilbert space). Thus in a toric configuration, the logical algebra condenses as $\bar{\mathfrak{su}}(4) \otimes \bar{\mathfrak{su}}(4) \xrightarrow{e^2 m^2 = \mathbb{I}} \bar{\mathfrak{su}}(2) \otimes \bar{\mathfrak{su}}(2)$, i.e. further error correcting qubits compared to a base $\mathbb{Z}_2$ Toric encoding. 

\emph{Error Correction and Reflection --}
To appreciate the rich dynamical processes at the interface of these two gapped phases we first examine error correction in the DS phase. Here errors excite both confined and deconfined anyons. The deconfined anyons of DS correspond to products with even overall topological charge. We define scattering processes conserving the good quantum numbers (equivalency classes modulo $e^2m^2$) at the boundary of two topological phases.

The quasiparticle types and locations are identified by syndrome extraction, e.g. with ancilla ququarts at the centers of order four vertex operators and qubits at plaquette centers to measure the order two plaquettes and short strings stabilizers. Error correction in the DS model consists of eliminating confined and deconfined particles as summarized in Algorithm ~\ref{alg:DS_QECC}. Further measurement errors partially spoil quasiparticle identification, so in practice stabilizers are measured several times before a highest-probability recovery map is applied\cite{Fowler2012}.

\begin{algorithm}
\caption{DS Error Correction} \label{alg:DS_QECC}
\begin{algorithmic}[1]
\LOOP
\STATE Measure  $\tilde{A},\tilde{B},\tilde{C}$  at each clock cycle
\IF{$\langle \tilde{C} \rangle =-1$} 
\STATE Examine $\langle \tilde{A} \rangle$ on adjacent vertices to deduce and undo confined $X,X^{\dagger},Z,Z^{\dagger}$ errors (or compositions thereof e.g. $X,\cdots,X$). $\tilde{B}$ is redundantly used as an $X-Z$ error type consistency check.
\ELSE { }
\STATE
$\tilde{A}_v$ and $\tilde{B}_p$ mark the deconfined excitations. 
\IF {$\langle \tilde{B}_p \rangle  = -1$}
\STATE $\exists$ a pair of $\tilde{A}_v$ diagonal to $B_p$ such that $\langle \tilde{A}_v \rangle = \pm i$. 
\IF {Both $\tilde{A}_v$s have the same phase}
\STATE Identify quasiparticle as $s$
\ELSE { }
\STATE Identify quasiparticle as $\bar{s}$.
\ENDIF
\ENDIF
\STATE Based on the excitation type and position, fuse pairs of $s$ (or $\bar{s}$) along \textit{contractible} loops moving them with the oriented short string operators\cite{Ellison2022}.
\IF {Only $\langle\tilde{A}_v\rangle = -1$}
\STATE $A_v$ marks a $s\bar{s}$ excitation. Fuse with its closest $s\bar{s}$ neighbor by application of $\prod_{j\in \gamma} Z^2_j$ along a \textit{contractible} loops.
\ENDIF
\ENDIF
\ENDLOOP
\end{algorithmic}

\end{algorithm}

\emph{Condensate-mediated scattering --}
\label{sec:AR}
We now discuss the admissible quasiparticle scattering processes at the boundary of the base $\mathbb{Z}_4$ and DS phases. 
Consider the application of the short string operator $\tilde{C}_h (\tilde{C}_v)$ along a vertical (horizontal) strip. Prior to condensation such a ribbon operators action would correspond to creating and moving a \textit{pair} of $e^2 m^2$'s, possibly around a handle for a logical $\bar{X}^2 \bar{Z}^2$ operation. 
Like the twists' proliferation of $e\times m = \epsilon$ excitations, \cite{Bombin2010} condensation proliferates $e^2 m^2$'s. The scattering processes in this work can be understood in analogy with the superconducting \textit{proximity effect}, whereby condensed Cooper pairs leak into an adjacent metal or insulator. Here $e^2m^2$s leak from the DS phase into the TC and can source logical errors.
This is illustrated in Fig~\ref{fig:condense_and_scatter} (c) where an $e^2 m^2$ pair is sourced into the surrounding TC. On the other hand $e, e^3, m, m^3, e m^2, e^3 m^2, e^2 m, e^2 m^3$ excitations correspond to \textit{confined} quasiparticles in DS regions. Technically, these quasiparticles can propagate into the DS phase but at an energetic penalty ($\langle \tilde{C}\rangle=-1$ along the path) scaling linearly with the penetration depth. This class of excitations, therefore, reflects off the DS condensate with high-probability as illustrated in panel (c). 

Fig~\ref{fig:condense_and_scatter} (d) illustrates the parity preserved quasiparticle conjugacy classes. Each of these corresponds to a distinct flavor of Andreev reflection and bound states in our superconductivity analogy\cite{Blonder1982,Beenakker1997,Law2009}. Stabilizer syndrome extraction allows microscopic measurement-based access to the particle types and their oscillations with respect to the condensed conjugacy classes. 

In practice, erroneous transitions \textit{between} conjugacy classes are inevitable and are quasi-particle poisoning events. One should begin with small un-condensed regions as it is simpler to track the particle parities and to reduce error rates (assuming it scales with the system size). Thus the lifetime of the topological oscillations of the quasiparticle and its $e^2 m^2$ conjugate bound state reflects the overall quality of an experimental realization. This high-level diagnostic lifetime can be directly compared with quasiparticle lifetime and decay rates across topological platforms\cite{Karzig2021}. 

\emph{Conclusions --} In this work we present and analyze the effects of twist-based topological deformations in $\mathbb{Z}_{N=2,4}$ gauge theories. To begin, we presented the twisted qubit code, which arises from a non-contractible twist. We also presented the twisted logical operators. The preparation and use of maximally entangled Bell states is a ubiquitous quantum computing task and the handled twist described here provides a simple scalability test towards Bell states for near-term topological codes. The twist doubles and triples the Hamming-distance of the logical $Y$ and $X$ errors respectively, i.e. with logical operators oriented perpendicular to the twist and traversing a cycle and co-cycle. The twist therefore provides a potentially useful dynamical strategy where a logical qubit is exchanged for enhanced protection in a strongly biased error channel. 

We then turned our attention to condensation of the $\mathbb{Z}_4$ model into a doubled-semion (DS) ordered condensate \cite{Iqbal2018, Ellison2022}. We examined the error correction properties and by condensing regions of various sizes we discuss condensation-mediated\cite{Kong2014} quasiparticle conjugacy classes. Scattering between these classes and the lifetimes of the oscillations is proposed as a topological metric. In this work we have only considered local twist constructions which could be generalized to non-local twist as a new type of wormhole \cite{Krishna2020}. We expect our described protocols and their generalizations to be useful in characterizing and understanding topological order in practice. During the preparation of this manuscript Refs.~\citenum{Barkeshli2022, Tantivasadakarn2022} appeared which discuss complementary directions.

\emph{Acknowledgements --}
This research was sponsored by the Quantum Science Center. The authors thank Lukasz Cincio, Shawn Cui, Gabor Halasz, and Travis Humble for helpful discussions. Tushar Pandey acknowledges the NSF-MSGRI program. 

\bibliography{main.bib}
\end{document}